\documentclass[conference]{IEEEtran}
\IEEEoverridecommandlockouts
\usepackage{cite}
\usepackage{amsmath,amssymb,amsfonts}
\usepackage{algorithmic}
\usepackage{graphicx}
\usepackage{textcomp}
\usepackage{xcolor}
\def\BibTeX{{\rm B\kern-.05em{\sc i\kern-.025em b}\kern-.08em
    T\kern-.1667em\lower.7ex\hbox{E}\kern-.125emX}}

\usepackage[numbers]{natbib} 
\usepackage{hyperref}

\usepackage{booktabs}
\usepackage[font=footnotesize,labelformat=simple]{subcaption}
\usepackage{color}
\usepackage{comment}
\usepackage{url}
\usepackage{bbm}
\usepackage[ruled,vlined]{algorithm2e}
\usepackage{algorithmic}
\usepackage{colortbl}

\newcommand{\hide}[1]{}
\newcommand{\cut}[1]{}

\usepackage{xcolor}
\newcommand{\covid}{COVID-19}

\makeatletter 
\def\@IEEEpubidpullup{8\baselineskip} 
\makeatother
\begin{document}

\IEEEoverridecommandlockouts
\IEEEpubid{
\parbox{\columnwidth}{\vspace{-4\baselineskip} Permission to make digital or hard copies of all or part of this work for personal or classroom use is granted without fee provided that copies are not made or distributed for profit or commercial advantage and that copies bear this notice and the full citation on the first page. Copyrights for components of this work owned by others than ACM must be honored. Abstracting with credit is permitted. To copy otherwise, or republish, to post on servers or to redistribute to lists, requires prior specific permission and/or a fee. Request permissions from \href{mailto:permissions@acm.org}{permissions@acm.org}.
\\ 
\small\textit{ASONAM '21}, November 8-11, 2021, Virtual Event, Netherlands \\ 
\copyright\space2021 Association for Computing Machinery. \\
ACM ISBN 978-1-4503-9128-3/21/11\ldots\$15.00 \\
\url{https://doi.org/10.1145/3487351.3488324}
\hfill}
\hspace{0.9\columnsep}\makebox[\columnwidth]{\hfill}}
\IEEEpubidadjcol

\title{Racism is a Virus: Anti-Asian Hate and Counterspeech in Social Media during the COVID-19 Crisis  }

\author{Bing He\textsuperscript{1}, Caleb Ziems\textsuperscript{1}, Sandeep Soni\textsuperscript{1}, Naren Ramakrishnan\textsuperscript{2},  Diyi Yang\textsuperscript{1}, Srijan Kumar\textsuperscript{1} 
\\~\textsuperscript{1} Georgia Institute of Technology, ~\textsuperscript{2} Virginia Tech 
\\~\textsuperscript{1}\texttt{\{bhe46, cziems, sandeepsoni, diyi.yang, srijan\}@gatech.edu},~\textsuperscript{2} \texttt{naren@cs.vt.edu}
\\}

\maketitle

\begin{abstract}
The spread of COVID-19 has sparked racism and hate on social media targeted towards Asian communities. 
However, little is known about how racial hate spreads during a pandemic and the role of counterspeech in mitigating this spread.
In this work, we study the evolution and spread of anti-Asian hate speech through the lens of Twitter. 
We create \texttt{COVID-HATE}, the largest dataset of anti-Asian hate and counterspeech spanning 14 months, containing over 206 million tweets, and a social network with over 127 million nodes. 
By creating a novel hand-labeled dataset of 3,355 tweets, we train a text classifier to identify hate and counterspeech tweets that achieves an average macro-F1 score of 0.832. 
Using this dataset, we conduct longitudinal analysis of tweets and users.
Analysis of the social network reveals that hateful and counterspeech users interact and engage extensively with one another, instead of living in isolated polarized communities.
We find that nodes were highly likely to become hateful after being exposed to hateful content.
Notably, counterspeech messages may discourage users from turning hateful, potentially suggesting a solution to curb hate on web and social media platforms.
Data and code is at  \texttt{\url{http://claws.cc.gatech.edu/covid}}.\footnote{The current paper is an extended version of the ASONAM 2021 paper with the same title.}
\end{abstract}
\section{Introduction}

The global outbreak of coronavirus disease 2019 or \covid{} caused widespread disruption in the personal, social, and economic lives of people. 
The upheaval 
has resulted in increased levels of fear, anxiety, and outbursts of strong emotions~\cite{montemurro2020emotional}. 
Hateful incidents throughout the world, such as acts of microaggression, physical and verbal abuse, and online harassment have increased during the pandemic~\cite{montemurro2020emotional}. 
Following the identified origin of \covid{} in China, racially motivated hate crime incidents have increasingly targeted the Chinese and the broader Asian communities, resulting in over 6,603 racially-motivated hateful incidents in the past year~\cite{nbcnews}.
Even the FBI warned of a potential surge in anti-Asian hate crimes motivated by \covid{} a year ago~\cite{fbiracism}.
The attacks in Atlanta, Georgia on March 16, 2021, which led to the death of six Asian women, show the grim reality of racial hate~\cite{nyt}.

\begin{figure}
\centering
        \includegraphics[width=\columnwidth]{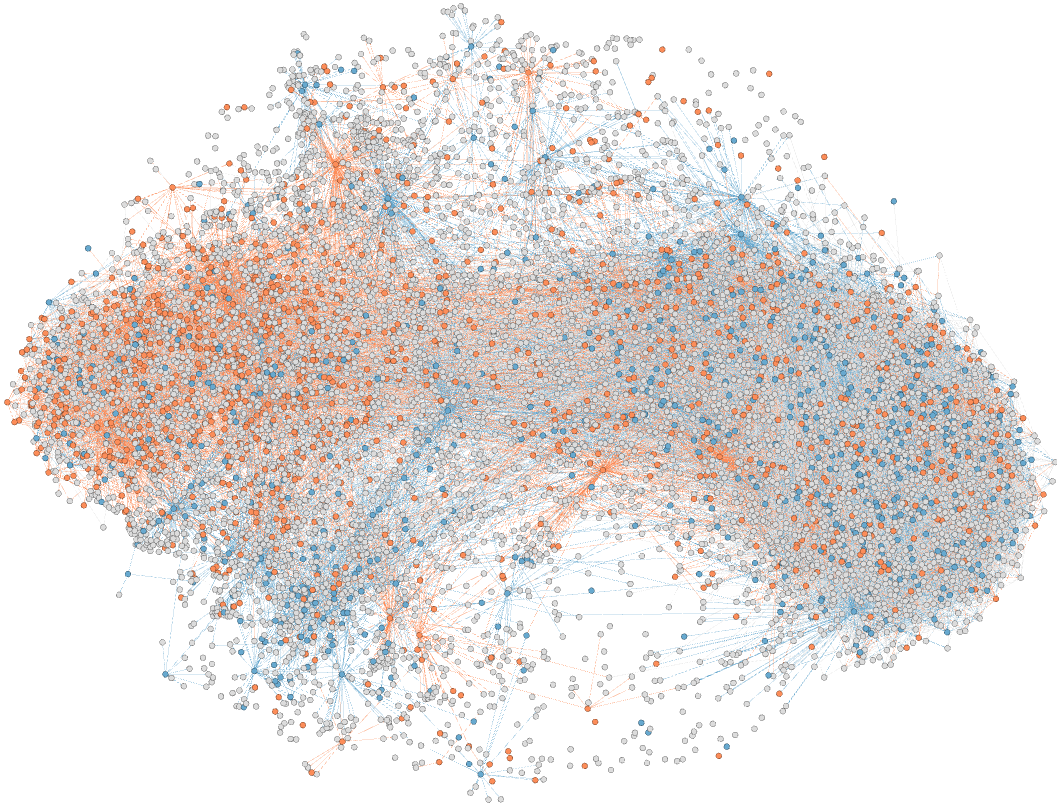}
    \caption{The \texttt{COVID-HATE} social network with hate nodes (orange), counterspeech nodes (blue), and neutral nodes (gray). 
    \label{fig:hate_network}
    }
    \vspace{-25pt}
\end{figure}

While there is mounting evidence of offline discriminatory acts and racism during \covid{}, the extent of such overtly hateful content on the web and social media is not widely known, especially their longitudinal pattern. 
Online hate speech has severe negative impact on the victims, often deteriorating their mental health~\cite{saha2019prevalence}. 
Prior research has shown that real-world crimes can be motivated by discussions on online social media platforms~\cite{relia2019race}. 
To counter hate speech, while efforts to educate about, curb, and counter hate have been made via social media campaigns (e.g. the \#RacismIsAVirus campaign), the success, effectiveness, and reach of counterspeech messages remain unclear. 
Thus, it is crucial to detect online hate speech to curb both online and physical harm, and monitor counterspeech messages to quantify their effectiveness, and inform future strategies to counter hate speech.

Recent research has been conducted on \covid{}-related hate and xenophobic online posts against Asians~\cite{vidgen2020detecting,schild2020go,vishwamitra2020analyzing, alshalan2020detection, al2021combating, lu2020fear}.
Building on these concurrent research works, we contribute several novel aspects to the understanding of this phenomenon. 
First,  we conduct a long-term longitudinal study of the hate and counterspeech ecosystem on Twitter to monitor the changes in social perception and stance towards the Asian community as the pandemic progressed. 
Second, we study the combined ecosystem of hate and counterspeech messages on Twitter, as opposed to studying them in isolation. This is important because both co-exist on the platform and influence each other simultaneously. Studying only one type of message (hate or counterspeech) is unable to uncover the influence they have on each other.

\textbf{Our contributions.} In this paper, we present \texttt{COVID-HATE}, the largest dataset of anti-Asian hate and counterspeech on Twitter in the context of the \covid{} pandemic, along with a 14 month-long longitudinal analysis of the Twittersphere. We make the following key contributions: 
\begin{itemize}
    \item We create a dataset of \covid{}-related tweets, containing over 206 million tweets made between January 15, 2020 and March 26, 2021. We also crawl the social network of users, containing over 127 million nodes and 910 million edges. 
    \item We annotate 3,355 tweets based on their hatefulness towards Asians as hate, counterspeech, or neutral tweets. We build a highly accurate text classifier to identify hate and counterspeech tweets, which achieves a macro-F1 score of 0.832. Using this classifier, we identify 1,227,116 hate and 1,154,289 counterspeech tweets in our dataset, making it the largest combined dataset of anti-Asian hate and counterspeech during \covid{}. A subgraph of user social network with nodes annotated with hate, counterhate, and neutral labels is shown in Figure~\ref{fig:hate_network}. 
    \item We conduct statistical, linguistic, and network analysis of tweets and users to reveal characteristic patterns of hate and counterspeech. We study social contagion in social network and find counterspeech tweets may lower the probability of neighboring nodes becoming hateful. 
\end{itemize}
\section{COVID-HATE: An Anti-Asian Hate and Counterspeech Dataset during \covid{}}

In this section, we describe \texttt{COVID-HATE}, a Twitter dataset containing COVID-19 anti-Asian hate and counterspeech tweets and social network. Table~\ref{tab:statistics} shows the data statistics.

{
\small
\begin{table}[t]
    \centering
    \resizebox{\columnwidth}{!}{%
    \begin{tabular}{l|l}
        \hline
        \textbf{Property} & \textbf{Statistic} \\\hline
        Duration & Jan 15, 2020--Mar 26, 2021 \\
        Number of tweets & 206,348,565 \\
        Number of (frac.) hateful tweets & 1,337,116 (0.64\%)\\
        Number of (frac.) counterspeech tweets & 1,154,289 (0.55\%)\\
        Number of (frac.) neutral tweets & 203,857,160 (98.81\%)\\
        \hline
        Number of users & 23,895,911\\
        Number of (frac.) hateful users & 697,098 (2.91\%) \\
        Number of (frac.) counterspeech users & 629,029 (2.63\%) \\
        Number of (frac.) neutral users & 22,477,616 (94.06\%) \\\hline
        Number of nodes in the social network & 127,831,666\\
        Number of edges in the social network & 910,630,334\\
        \hline
    \end{tabular}
    }
    \caption{Statistics of \texttt{COVID-HATE} dataset, containing anti-Asian hate and counterspeech tweets and social network in the context of \covid{}.
    \label{tab:statistics}
    }
\end{table}
}

\subsection{Tweet Dataset}
\label{sec:twitter_data}

We adopted a keyword-based approach to collect relevant \covid{} tweets through Twitter's official APIs.
Specifically, we used a collection of keywords and hashtags belonging to three sets: 
(a) \texttt{covid-19 keywords} are terms referring to \covid{} which are used to collect tweets related to the pandemic, 
(b) \texttt{hate keywords} are keywords and hashtags indicating anti-Asian hate amidst \covid{}. To compile this list, we first took the hate keywords from existing papers~\cite{chen2020covid} and news articles~\cite{dylan2020}. 
We then expanded this list by including co-occurring hate hashtags observed in an initial tweet crawl. 
We also included Asian slurs listed in Hatebase\footnote{https://hatebase.org/}.
Finally, (c) \texttt{counterspeech keywords} are keywords and hashtags that were used to organize efforts to counter hate speech and support Asians. 
These keywords were listed in news articles covering counterhate efforts during the initial phases of the data collection setup~\cite{abc,businessinsider,prweek}. In total, we used 42 keywords as shown in Table~\ref{tab:keywords}. 
During the process, we intentionally created a broad list of keywords to ensure high recall. This may result in collection of borderline-relevant tweets as well, which can later be identified and removed in the filtering step via a classifier, which we describe later. 

We utilized a combination of Twitter’s Streaming API and Twitter's Search API to collect the data. In particular, we started collecting real-time tweets from March 28, 2020 using the Streaming API. We also used the Search API to get historical tweets containing the aforementioned keywords since January 15, 2020, as there were no existing datasets containing hate and counterspeech discussions spanning the initial months of the pandemic (January to March 2020). 

Using these keywords, we collected 206,348,565 English-language tweets made by 23,895,911 users between January 15, 2020 and March 26, 2021. This data does not contain retweets so that the analysis is focused on original content. Since keyword-based collection can return tweets that are not relevant to the topic of study, we develop a classifier to remove irrelevant tweets (as explained in the next subsection).

{
\small
\begin{table}[t]
    \centering
     \resizebox{\columnwidth}{!}{%
    \begin{tabular}{l|l}
    \hline
    Category & Keywords \\\hline
       \covid{}  &  coronavirus, covid 19,
        covid-19, covid19, corona virus\\
     Hate & \#CCPVirus, \#ChinaDidThis, \#ChinaLiedPeopleDied, \\
     keywords & \#ChinaVirus, \#ChineseVirus, chinese virus, \\ 
     & \#ChineseBioterrorism, \#FuckChina, \#KungFlu, \\
     & \#MakeChinaPay, \#wuhanflu, \#wuhanvirus, wuhan virus, \\
     & chink, chinky, chonky, churka, cina, cokin,  \\
     & communistvirus, coolie, dink, niakou\'e, pastel de flango, \\
     & slant, slant eye, slopehead, ting tong, yokel \\
     Counterspeech & \#IAmNotAVirus, \#WashTheHate, \#RacismIsAVirus, \\
     keywords & \#IAmNotCovid19, \#BeCool2Asians, \#StopAAPIHate,\\
     & \#ActToChange, \#HateIsAVirus\\
     \hline
    \end{tabular}
    }
    \caption{The list of keywords and hashtags used for comprehensive data collection. 
    }
    \label{tab:keywords}
     \vspace{-15pt}
\end{table}
}

\textbf{Twitter Network Construction:}
In addition to the tweets, we crawled the ego-network (i.e., the followers and followees) of a randomly-sampled subset of users who made at least one COVID-19 tweet. 
Due to Twitter's GET API limitations, we obtain the latest maximum 5,000 followers and 5000 followees per user. 
A total of 489,011 users' neighborhoods were crawled. 
The resulting network has 127,831,666 nodes and 910,630,334 edges.

\subsection{Annotating Anti-Asian \covid{} Hate and Counterspeech}
To identify tweets relevant to our study of hate and counterspeech, we hand-label a subset of tweets and create a textual classifier to label the rest. 
Even though tweets may have explicitly hateful hashtags, categorizing tweets simply based on the presence (or absence) of a hashtag and keyword is insufficient because hashtags can be added to gain visibility and promote tweets. 
Conversely, a tweet can be hateful even without having a hateful hashtag.
The same is true for counterspeech tweets. 
Thus, we developed a rigorous annotation process to establish the ground truth categories of tweets based on the tweet content. 

Please note that Vidgen et al.~\cite{vidgen2020detecting} has concurrently created an annotated dataset with hate and counterspeech labels.
However, there are only 116 counterspeech tweets in their annotations, which is usually insufficient to train a machine learning model that is representative of the class.
To avoid this issue, we manually labelled and created a new and more balanced \covid{} data with hate and counterspeech labels. Since the work by Vidgen et al.~\cite{vidgen2020detecting} was done concurrently, our future research can consider integrating the two set of annotations to create a more comprehensive dataset. Below we describe our annotation process.

We labeled the tweets into the following three broad categories, as we define below.

\textbf{Anti-Asian \covid{} Hate Tweets:}
We build on previous studies of racial hate from the social and information science literature to define anti-Asian hate. 
Specifically, Parekh et al.~\cite{parekh2012there} and Fortuna et al.~\cite{fortuna2018survey} established that hate speech is directed at an individual or group based on ``an arbitrary or normatively irrelevant feature,'' and that it casts the target as an ``undesirable presence and a legitimate object of hostility''. More recent work on pandemic-era prejudice and anti-Asian hate crimes \cite{roberto2020stigmatization,reny2020xenophobia,gover2020anti} understand these behaviors as grounded in ``\textit{othering}'' attitudes, i.e., isolate the outgroup like an infectious disease \cite{kam2019infectious}.
Building on this, we define anti-Asian \covid{} hate as antagonistic speech that: (a) is in the context of \covid{}, (b) is directed towards an Asian entity (individual person, organization, or country), and (c) \textit{others} the Asian outgroup through intentional opposition or hostility
by verbal abuse, derogatory language (racism, disrespect), or blame for causing, spreading, being responsible for, or hiding \covid{}. 
One overt example of anti-Asian hate we considered is (censorship ours):
\begin{quote}{\small
\textit{F*ck Chinese scums of the Earth disgusting pieces of sh*t learn how to not kill off your whole population of pigs, chickens, and humans. coronavirus, \#wuhanflu, \#ccp, \#africaswine, \#pigs, \#chickenflu nasty nasty China clean your f*****g country.}
}
\end{quote}
In our definition, we distinguish hate from strong criticism of Asian entities by using this \textit{othering} language. For example, ``\textit{China let the pandemic get out of hand}'' is not hate speech, but ``\textit{It is just like the Chinese to let the pandemic get out of hand}'' is hate, precisely because it typifies Chinese people as a distinct \textit{other} -- a class of people who are fundamentally more likely to instigate a pandemic. In this work, we do not consider the motivation or reason behind the hate speech (e.g., whether it is linked to a conspiracy theory), as it is beyond the scope of the current work.

\textbf{\covid{} Counterspeech Tweets:}
This category of tweets is also in the context of \covid{} and can either:
(a) explicitly identify, call out, criticize, condemn, challenge, or oppose racism, hate, or violence towards an Asian entity or 
(b) explicitly support, express solidarity towards, or defend an Asian entity. These tweets can either be direct replies to hateful tweets or be stand-alone tweets, but they must be explicit. An example of a tweet in this category is as follows: 
\begin{quote}{\small
\textit{The virus did inherently come from China but you can’t just call it the Chinese virus because that’s racist. or KungFlu because 1. It’s not a f*****g flu it is a Coronavirus which is a type of virus. And 2. That’s also racist.}}
\end{quote}
The intensity of this example shows why counterspeech detection is challenging: an ignorant keyword-matching system might conflate the profanity here with hate speech. This is a direct reply message, but counterspeech can also include generalizations like ``\textit{The treatment of Asians specifically, during this time, is UNACCEPTABLE}.'' In both cases, the counterspeech is explicit. We do \textit{not} consider more implicit statements of solidarity to be counterspeech, like the rhetorical question ``\textit{Why do we have to be so divisive and cruel about this pandemic?}'' where the author does not explicitly specify who the target of cruelty is.

\textbf{Neutral and Irrelevant Tweets:}
These tweets neither explicitly nor implicitly convey hate, nor counterspeech, but contain keywords related to \covid{}. Tweets in this category also include news, advertisements, or outright spam. One example of a tweet in this category is:
\begin{quote}{\small
\textit{COVID-19: \#WhiteHouse Asks Congress For \$2.5 Bn To Fight \#Coronavirus: Reports \#worldpowers \#climatesecurity \#disobedientdss \#senate \#politics \#news \#unsc \#breaking \#breakingnews \#wuhan \#wuhanvirus https://t.co/XipNDc}}
\end{quote}

\textbf{Annotation process:}
We trained two undergraduate annotators to recognize anti-Asian \covid{} hate tweets, \covid{} counterspeech tweets, and neutral/irrelevant tweets using the above definitions. Both annotators are of Asian descent (one Chinese and one Indian). One co-author supervised the annotation process. After practicing on a set of 100 tweets and discussing disagreements with the supervising co-author,
the annotators each independently labeled the same set of 3,255 tweets, which were randomly sampled from the collected dataset. Since the majority of tweets were expected to be neutral, we over-sampled tweets that contained anti-Asian hate, and counterspeech terms. This ensured our labeling process yielded sufficient hate and counterspeech tweets to train a classifier. The annotation process took six weeks. 

The two annotators agreed on 68\% of the data, with Cohen's Kappa score of 0.448 for hate and 0.590 for counterspeech, indicating a moderate inter-rater agreement that is typical of hate speech annotation \cite{ross2017measuring, vidgen2020detecting}. We removed the tweets where the two annotators disagreed and were left with 429 hate, 517 counterspeech, and 1,344 neutral tweets. The annotators also identified 110 tweets containing hatefulness or aggression towards non-Asian groups. Since our goal is to study anti-Asian \covid{} hate, we drop the latter set of tweets. We focus only on anti-Asian hate, counterspeech, and neutral tweets in the remainder of this paper.

\subsection{Anti-Asian Hate and Counterspeech Text Classifier}
\label{sec:classifier_dev}

We use the annotated tweets to train a text-based machine learning classifier to label tweets as anti-Asian hate, counterspeech, or neutral. We create the following three sets of features that are used for classification.

\begin{table}[t!]
    \centering
\begin{tabular}{lrrr}
\hline
{Feature set} &  Precision &  Recall &  F1 score \\
\multicolumn{4}{c}{\textbf{Anti-Asian hate tweet detection}}\\
\hline
Linguistic          &      0.541 &   0.233 &     0.323 \\
Hashtag             &      0.100 &   0.002 &     0.005 \\
BERT                &      0.765 &   0.760 &     0.762 \\

\multicolumn{4}{c}{\textbf{Counterspeech tweet detection}}\\\hline
Linguistic          &      0.483 &   0.189 &     0.267 \\
Hashtag             &      0.800 &   0.029 &     0.056 \\
BERT                &      0.839 &   0.868 &     0.853 \\
\multicolumn{4}{c}{\textbf{Neutral tweet detection}}\\\hline
Linguistic          &      0.632 &   0.891 &     0.739 \\
Hashtag             &      0.591 &   0.999 &     0.743 \\
BERT                &      0.886 &   0.874 &     0.880 \\
\hline
\end{tabular}
    \caption{Tweet classification performance of different feature sets with a neural network classifier. The BERT model has the best classification performance in all three tasks.}
    \label{tab:hate_model_performance}
     \vspace{-15pt}
\end{table}

\noindent $\bullet$ \textbf{Linguistic Features.}
This set contains a total of 90 features spanning  stylistic, metadata, and  psycholinguistic categories, together representing the linguistic properties of the text.
These features have previously been very effective in identifying hate speech and cyberbullying~\cite{fortuna2018survey}.

\noindent $\bullet$ \textbf{Hashtag features.}
These features create a vector representation of each tweet representing the number of occurrences of each hashtag and keyword listed in Table~\ref{tab:keywords}.
The presence of keywords can be potential indicators of the tweet category. 
For example, a tweet containing `\#RacismIsAVirus' is more likely to be a counterspeech tweet than a hate tweet. 
The hashtag feature is 42-dimensional, representing the number of hashtags and keywords used for collection.

\noindent $\bullet$ \textbf{Bert Tweet Embeddings.}
The above two approaches extract granular features from tweets, but can ignore semantic meaning of the tweet. 
Thus, to incorporate the word-level and sentence-level semantics, we embed each tweet using the BERT base uncased text embedding model~\cite{devlin2018bert} and it gives an embedding representation of each tweet. 
Next, the embedding is used as an input to a neural network classifier with one feed-forward layer. Finally, the BERT classification model is fine-tuned as fine tuning provides superior classification performance~\cite{devlin2018bert}.

\textbf{Model training.}
Similar to the BERT classifier, one-layer feed-forward neural network classifiers are trained using linguistic features and hashtag features. 
After extensive grid search, we set all neural network parameters for batch size, number of epochs, and learning rate as 8, 3 and 1$e^{-5}$, respectively, to obtain the best classification results.
We conducted five-fold cross-validation on the hand-annotated dataset. 

Performance of the models was measured using precision, recall and F1 scores, as shown in Table~\ref{tab:hate_model_performance}.
In the table, we note that BERT embedding features perform better than linguistic and hashtag features by a significant margin. 
Hashtag features typically are the worst, as they convey low semantic meaning. 
We had also created ensemble methods by combining all three set of features together and training a neural network model, however, the model performance was similar to that of the BERT model. 

Overall, the BERT model has a very high precision and recall in categorizing tweets. Moreover, using the BERT classifier prevents the model from being overly reliant on the presence of specific hashtags and keywords, whose popularity may change over time. 
Thus, we use the BERT model trained on all the annotated data to label the rest of the tweets in the dataset. This results in about 1.337 million hate tweets and 1.154 million counterspeech tweets, which we analyze in the rest of the paper. 
\section{Longitudinal Characterization of\\ \covid{} Hate and Counterspeech}

In this section, we use the \texttt{COVID-HATE} dataset to analyze the patterns of hate and counterspeech in the Twitter ecosystem. We focus our analysis on the evolution and spread of hate and counterspeech and the characteristics of the users. 
To characterize the temporal changes in trends, we will compare the statistics between the year 2020 (from January 15, 2020 to December 31, 2020) and the year 2021 (from January 1, 2021 to March 26, 2021).

\subsection{The Ebb and Flow of Hate and Counterspeech}

Here we consider the longitudinal spread of hate and counterspeech tweets in the Twitter ecosystem. 

\begin{figure}
    \centering
    \includegraphics[width=0.8\columnwidth]{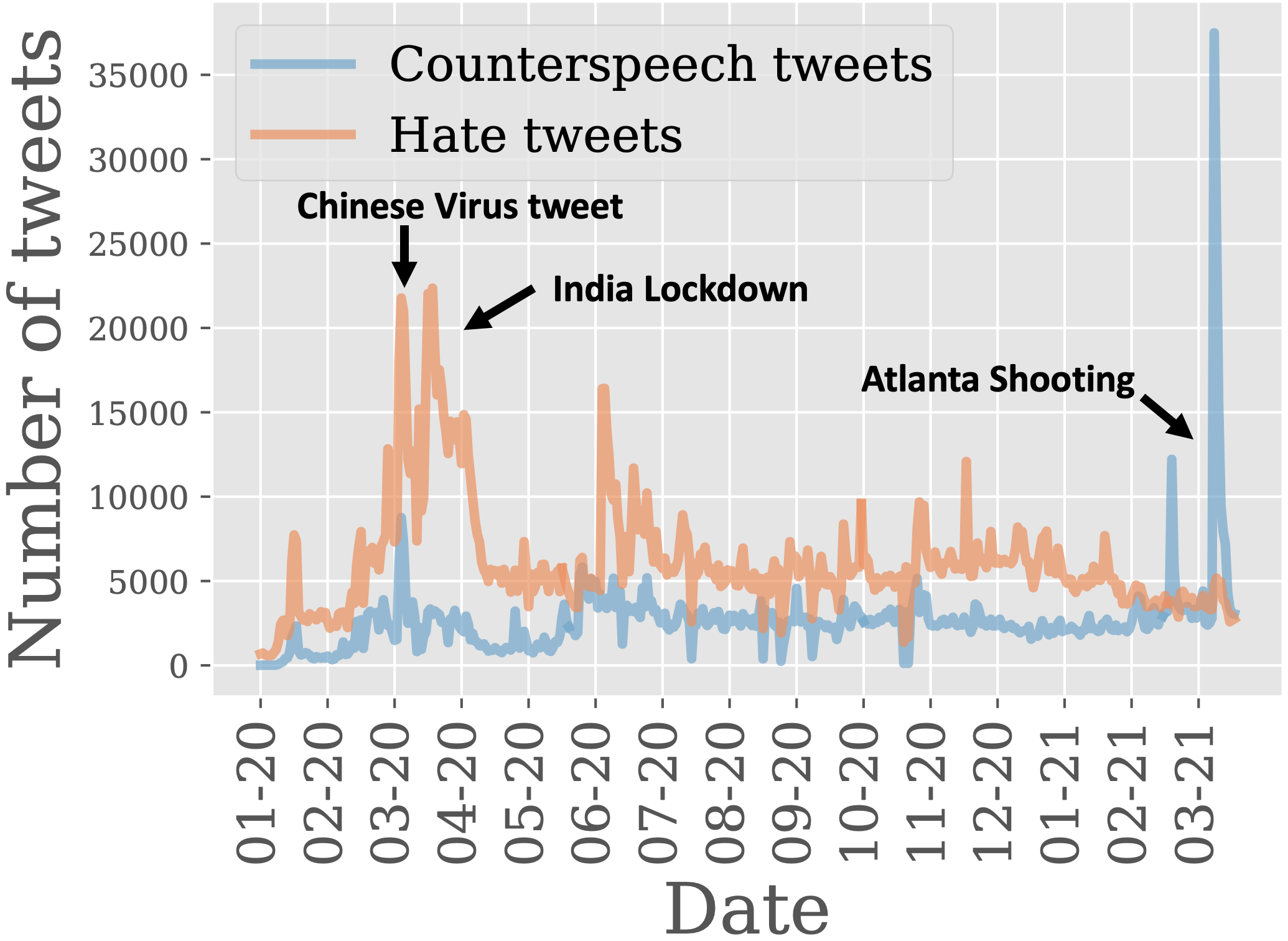}
    \vspace{-5pt}
    \caption{The number of hate and counterspeech tweets from January 15, 2020--March 26, 2021. 
    \label{img:hateful_tweets_count}
    }
    \vspace{-15pt}
\end{figure}

\textbf{Hate tweets were more frequent than counterspeech tweets in the year 2020.}
Figure~\ref{img:hateful_tweets_count} shows the daily distribution of hate and counterspeech tweets.
First, we note that hate tweets outnumber counterspeech tweets throughout the timeline during 2020. Next, the number of hate and counterspeech tweets was negligible-to-low during the early phases of the pandemic in January, 2020 and February, 2020.
We observe the spike in hate speech between March 16, 2020 and March 19, 2020. 
There were several major spikes in daily hate tweets throughout 2020, which exceeded the daily counterspeech tweets.

\textbf{Counterspeech tweets increased dramatically after the 2021 Atlanta shooting.} Counterspeech messages typically had lower volume throughout 2020 compared to hate tweets. However, after the Atlanta Spa shooting on March 16, 2021~\cite{nyt}, there was a dramatic increase in the number of counterspeech tweets in March, 2021. Counterspeech tweets increased by 401.2\% within one week, while we observed that hateful tweets also surprisingly rose by 17.9\%. The spike in counterspeech signals the Twittersphere expressing sympathy and solidarity towards the Asian community.

Please note that even though the keywords and hashtags used for data collection were selected during the early phase of the pandemic (March 2020), the dataset reveals spikes in hate and counterspeech throughout the 14 month period.

\begin{figure}[t]
    \centering
    \vspace{-3mm}
    \includegraphics[width=0.65\columnwidth]{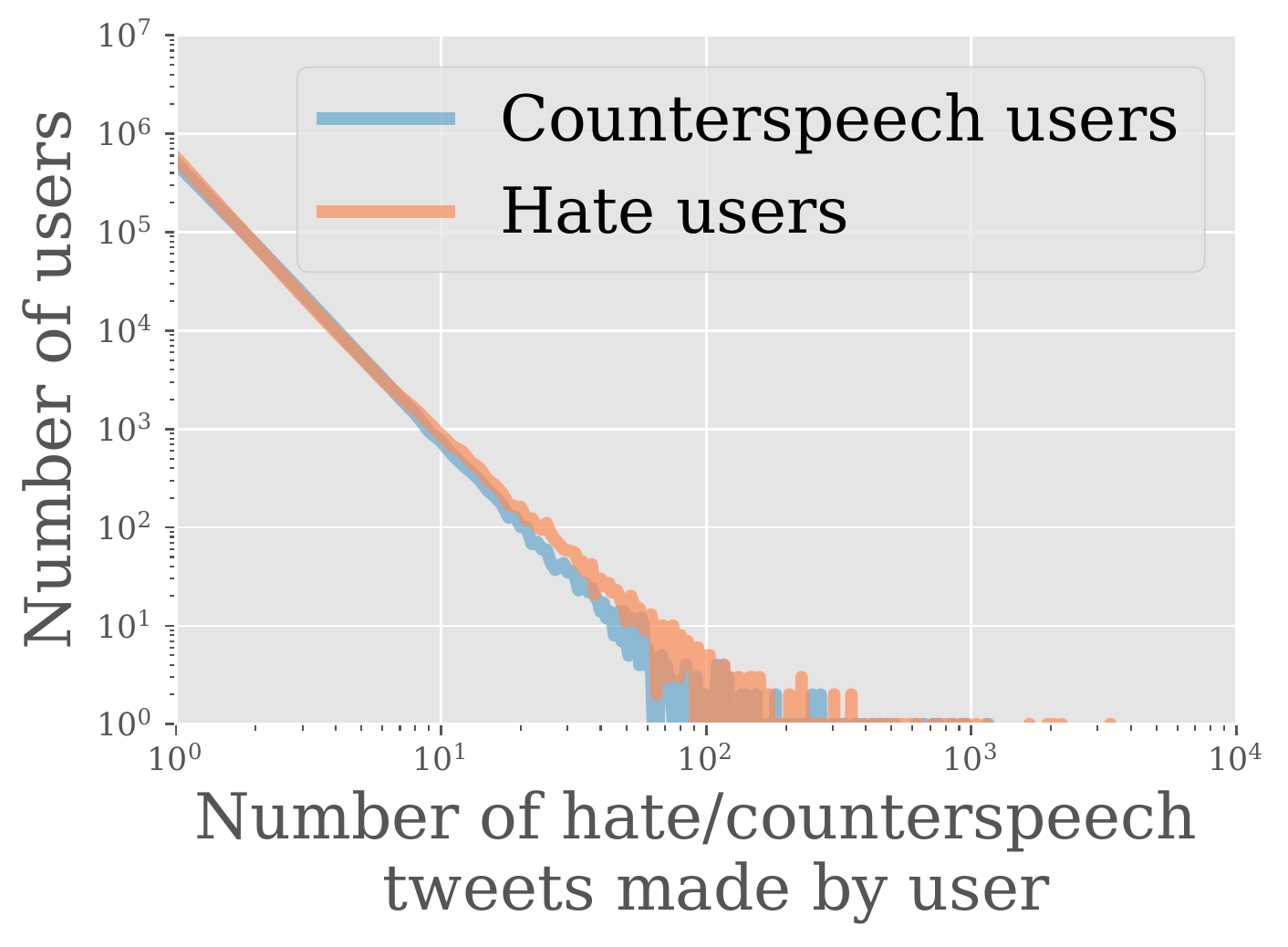}
    \vspace{-5pt}
    \caption{Distribution of the number of hate and counterspeech tweets made by users shows a long tail pattern. 
    }
    \label{fig:hate_activity}
     \vspace{-5pt}
\end{figure}

\subsection{User Activity and Interaction Behavior}
We analyze the properties of the users who produce hate and counterspeech tweets.
Following the tweet categorization labels, we categorize users, based on their tweets, into one of the following: \textit{hate}, \textit{counterspeech}, \textit{dual}, or \textit{neutral}
Hate users make at least one hate tweet but no counterspeech tweets. Similarly, counterspeech users make at least one counterspeech tweet but no hate tweets. 
Users who tweet from both categories are categorized as dual users. Finally, users who make at least one \covid{} tweet (and thus, are part of our dataset), but no hate or counterspeech tweets, are labeled as neutral. 
Among the 23,895,911 users in the dataset, most of the users (94.06\%) are neutral, 697,098 (2.92\%) are hateful, 629,029 (2.63\%) are counterspeech users, and a very small fraction of users (0.39\%) are dual. This distribution mimics the category-wise tweet distribution. Our following analysis focuses on hate, counterspeech, and neutral user categories. Due to low volume, we do not emphasize on the dual users, which can be worth exploring in future studies.

Figure~\ref{fig:hate_activity} shows the distribution of the number of hate tweets (counterspeech tweets, respectively) made by hate users (counterspeech users, respectively). We observe that both distributions exhibit a long tail, showing that most users make few relevant tweets and only a handful of users are responsible for spreading most of the hate propaganda and counterspeech messages.

\textbf{User activity before activation.}
When a user sends his or her first hate or counterspeech tweet, we say he or she becomes `activated.' 
Recall that hate users never make a counterspeech tweet and similarly, counterspeech users never make a hate tweet. 
Here we evaluate the pre-activation behavior of hate user and counterspeech user with respect to \covid{} discussions. 

First, we find that before activation, hateful users make fewer \covid{}-related tweets (an average of 14.28 COVID-related tweets) compared to counterspeech users (an average of 28.62 COVID-related tweets). 
This difference in \covid{} engagement is statistically significant as measured by Mann-Whitney U test ($p < 0.001$).\footnote{Unpaired t-test values are computed via Mann-Whitney U test.}

We also compared the sentiment and psycholinguistic properties of hate users and counterspeech users, prior to activation. 
On average, hate users wrote shorter tweets (120.89 vs. 160.38 characters; $p<0.001$), used fewer URLs (0.413 vs. 0.528; $p < 0.001$) and tagged others less often (0.775 vs. 0.918; $p<0.001$). 
Additionally, their overall sentiment scores were also more neutral (0.792 vs. 0.781 score; $p<0.001$). 
These statistics indicated that hate users made fewer and more neutral in \covid{} discussions compared to counterspeech users, prior to activation.
\textbf{User activity after activation.}
Here we contrast the post-activation behavior of users with their pre-activation behavior. 
We find that after activation, both hate and counterspeech users make more \covid{}-related tweets (20.78 and 49.66 COVID-related tweets on average, respectively; $p < 0.001$) and longer tweets (21.32 vs. 28.41 words per tweet; $p < 0.001$). Similar to pre-activation, counterspeech users make more \covid{}-related tweets than hate users.

\subsection{Social Network Connectivity Structure}
In this section, we examine the user-user social connectivity in the hate and counterspeech ecosystem. 
As described in the dataset section, we crawled the social network containing over 127 million nodes and 910 million edges. 
Out of these, 1,380,613 nodes have made at least one \covid{}-related tweet. The rest of the nodes are part of the network as they are neighbors of these nodes.
Figure~\ref{fig:hate_network} shows a subgraph of this network, with nodes colored according to their category (hate, counterspeech, or neutral).

To understand the differences in how hate and counterspeech users behave, we compare their ego-networks. 
We find that on average, counterspeech users are better connected than hate users---counterspeech users follow more users compared to hate users (1201.84 vs. 828.40; $p<0.001$) and are followed more by other users (1249.42 vs. 759.96; $p < 0.001$).

\begin{figure}
\centering
        \includegraphics[width=0.48\columnwidth]{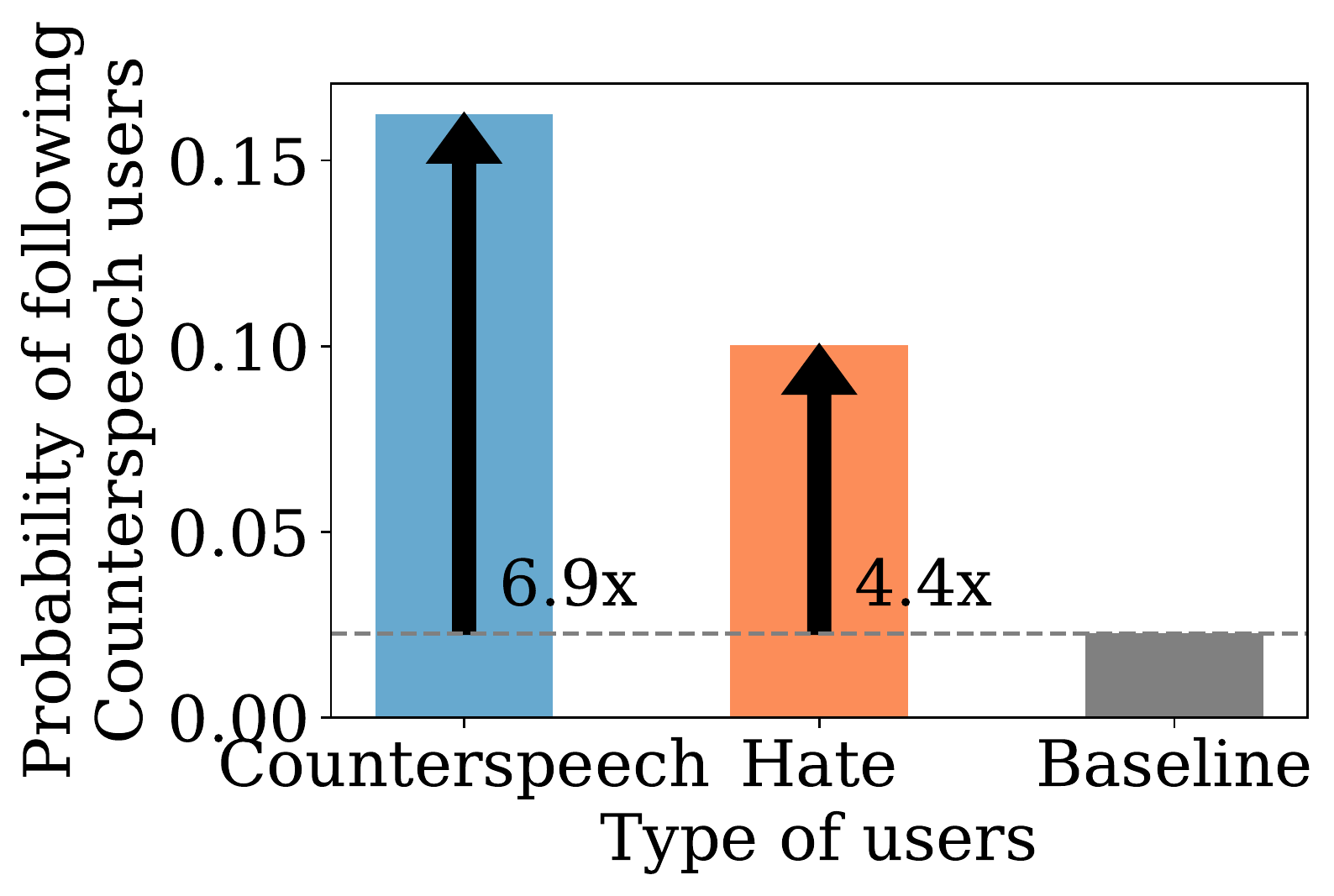}
        \includegraphics[width=0.48\columnwidth]{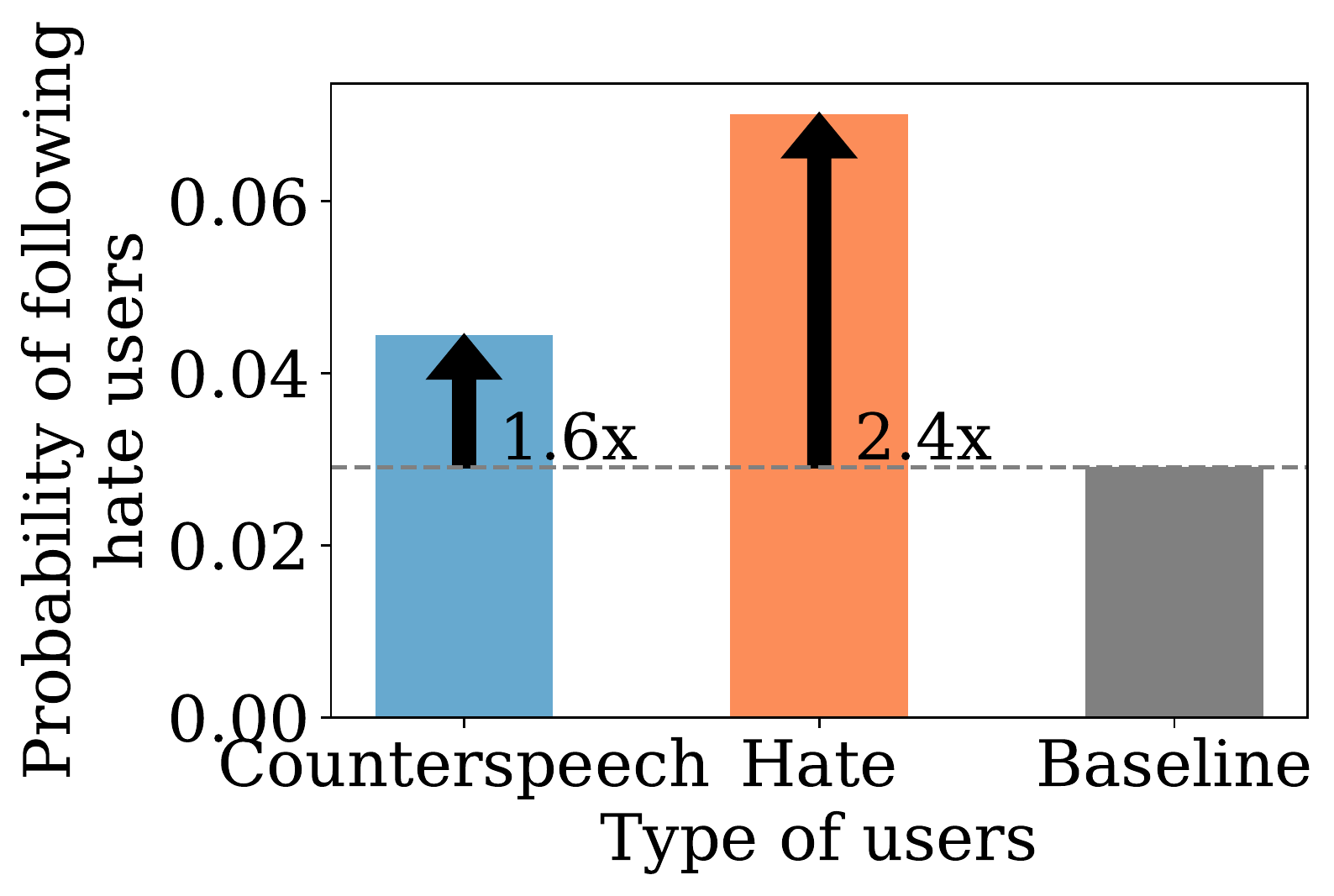} 
        \vspace{-5pt}
    \caption{Social network of hate and counterspeech users: Hate and counterspeech users are highly interconnected and exhibit homophily.
    \label{fig:network_baseline}
    }
    \vspace{-10pt}
\end{figure}

\noindent \textbf{Intragroup and intergroup connectivity.}
We analyze the connectivity of users within and across the different groups to establish if nodes express homophily or form echo chambers. 
Simply comparing their probability of creating edges to nodes of a certain group is not sufficient as it is confounded by the node degrees and node distribution across categories. 
Thus, we create a network baseline to model the expected behavior of nodes and compare against this baseline~\cite{leskovec2010signed}. 

Specifically, the baseline networks are created by randomly shuffling the edges, while keeping the set of nodes the same. 
The node degrees are preserved and each node is ensured to have the same number of \covid{} neighbors as it did in the original network, though the neighbors change during shuffling. 
We compute the aggregate ego-network statistics across several versions of baseline networks (100 in our experiments).

We compare the observed and the baseline behavior using the probability of connecting to hate, counterspeech, and neutral nodes. 
Figure~\ref{fig:network_baseline} presents the results.

\textbf{Nodes exhibit homophily.} 
First, we examine the propensity for hate and counterspeech nodes to connect with nodes within their own group. 
In Figure~\ref{fig:network_baseline} (left), we show that counterspeech users are 6.92$\times$ more likely to connect to other counterspeech users compared to the baseline behavior. 
Similarly, the right figure shows that hateful users connect with other hateful users 2.42$\times$ more than expectation. 
Thus, nodes are preferentially connected to other nodes in the same group.

\textbf{Do hateful and counterspeech users form polarized communities?}
Echo chambers and polarization are commonly-observed phenomena in social media, which are responsible for the spread of propaganda and misinformation~\cite{del2016echo}. 
However, it is not known whether echo-chambers exist in the hate network as well. 
Given that nodes preferentially connect to similar nodes, four scenarios are possible. 
(1) Hate and counterspeech users live in isolated echo-chambers, where these groups do not interact with one another. 
(2) On the other extreme, the two groups interact highly with each other, possibly exhibiting conflict. 
The remaining two possibilities are that the out-group connections are one-sided. 

Figure~\ref{fig:network_baseline} illustrates the empirically-observed behavior.
Both hate and counterspeech nodes are more likely to connect with one another than expected. 
Precisely, hateful users follow counterspeech users 4.45$\times$ more than expected (left figure) and counterspeech users are 1.62$\times$ more likely to follow hateful users compared to the baseline (right figure).

Altogether, these indicate that hateful and counterspeech users are highly engaged and closely interact with each other. 
\vspace{-1mm}
\section{Influence of Counterspeech on  the Spread of Hate}

Antisocial behavior, such as hate speech, abuse, and trolling, have been shown to exhibit social contagion ~\cite{mathew2019spread}. 
However, whether hate speech by a user motivates its neighbors to be hateful during the \covid{} pandemic is unknown.
Moreover, how the spread of counterspeech messages impacts the spread of hate messages is also not known, which is important to infer and quantify the role of counterspeech in curbing the spread of hate, if any.
Thus, we investigate the within-group and across-group influence on the diffusion of hate messages.

We quantify influence as the likelihood of a user to become hateful (i.e., writing an anti-Asian hate tweet for the first time) after a user is exposed to any number of hate or counterspeech tweets from his or her neighbors. 
We refer to a user's change of state from the neutral state to hate/counterspeech state after observing neighbors' messages as an \textit{infection}.
We model the dynamics of hate/counterspeech infection as an event cascade. 
The cascade is a temporally-ordered sequence of events of the nodes that transition from neutral to hate or counterspeech states. 
Each cascade is associated with a function $Risk_{s \rightarrow s'}(n)$ that quantifies the probability that a user transitions from neutral to category $s' \in \{hate, counterspeech\}$ after $n$ neighbors have become part of category $s \in \{hate, counterspeech\}$.
Neighbors are obtained from the social network. 
The infection risk function is calculated as: 
\begin{equation}
    Risk_{s \rightarrow s'}(n) = \frac{|Infected_{s'} \cap Exposed_{s}(n)|}{|Exposed_{s}(n)|}
\end{equation}
where $Infected_{s'}$ is the set of users already infected with type $s'$ and $Exposed_{s}(n)$ is the set of users with at least $n$ neighbors of type $s$.

\begin{figure}
    \begin{subfigure}[t]{0.49\columnwidth}
        \centering
       \includegraphics[width=\textwidth]{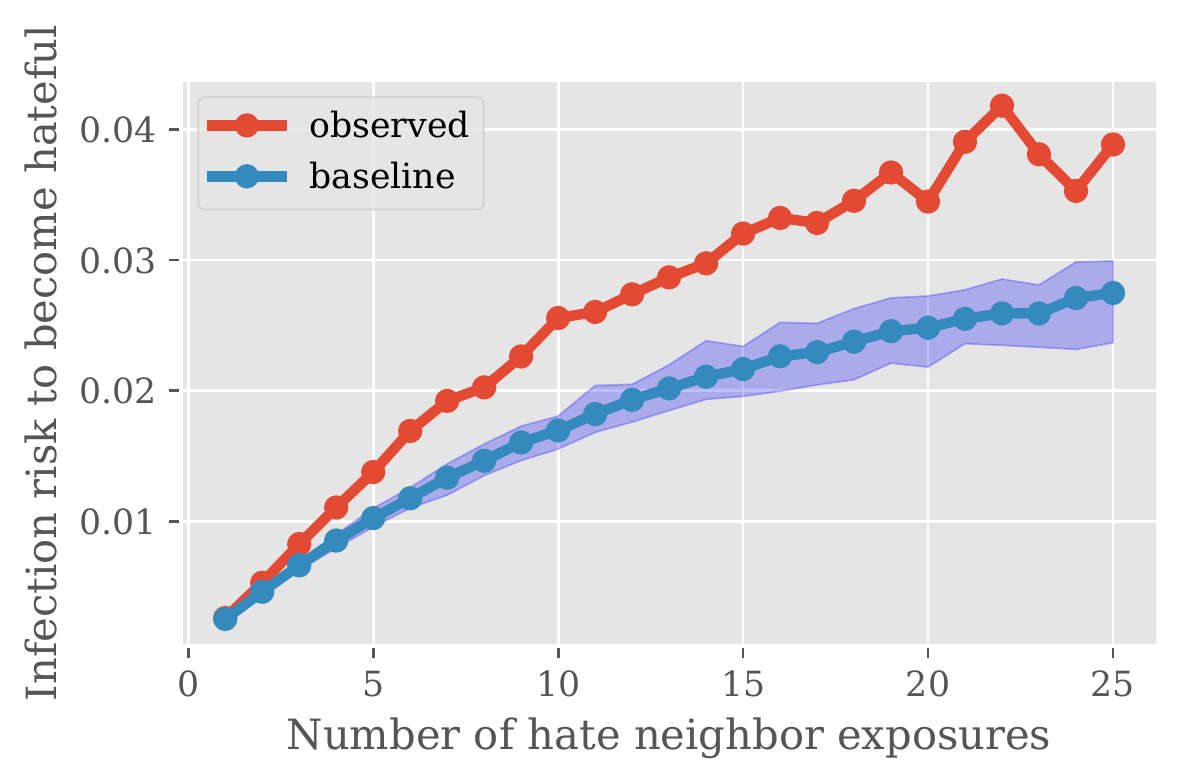}
        \caption{Hate $\rightarrow$ hate  \label{fig:hate2hate}}
    \end{subfigure}
    \begin{subfigure}[t]{0.49\columnwidth}   
        \centering 
        \includegraphics[width=\textwidth]{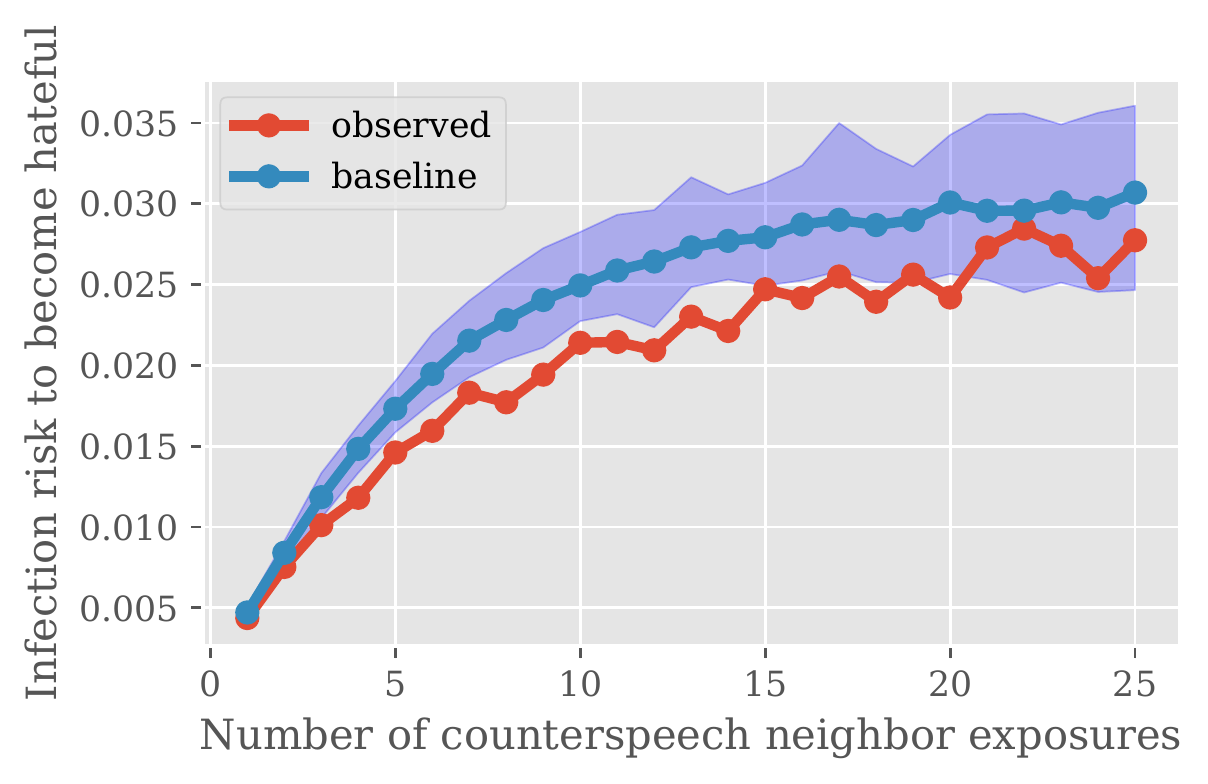}
        \caption{Counterspeech $\rightarrow$ hate  \label{fig:anti2hate}}
    \end{subfigure}

    \caption{The impact of hate speech and counterspeech on the spread of hate. 
    \label{fig:social-contagion-2020}
    }
    \vspace{-10pt}

\end{figure}

The infection risk in a network is conflated not only by users' influence on one another, but also by homophily---the tendency of similar users to cluster in the network. We have already shown in the previous sections that hate and counterspeech users exhibit homophily. To tease out the effect of influence from homophily, we create a null model that measures the baseline risk of infection solely due to homophily, without any user-to-user influence.
We follow the technique by \cite{anagnostopoulos2008influence} for this analysis. 
We randomly shuffle the order of cascade events and calculate the infection risk in the random cascade.
The social network remains fixed. 
We compare the mean baseline infection risk observed across 100 shuffled cascades to the empirically observed infection risk. 
If the empirical infection risk exceeds the baseline risk, then social contagion is responsible for the spread of infection (hate or counterspeech). On the other hand, if the empirical infection risk is lower than the baseline risk, then social contagion inhibits the spread of infection. 
The results are shown in Figure~\ref{fig:social-contagion-2020}. 

Figure~\ref{fig:hate2hate} shows that exposure to hate speech increased the likelihood of adopting hate speech, compared to the baseline.
Moreover, the likelihood of hate adoption increased with the number of exposures. The effect of counterspeech on hate speech contagion is crucial to investigate as it can shed light on potential counter-measures. 
As we can see, counterspeech might deter the spread of hate speech compared to the baseline, as shown in Figure~\ref{fig:anti2hate}, thus potentially showing low social inhibition effect. The change over the years might show a positive trend towards counterspeech mitigating hate speech, potentially influenced also by external societal factors.

\section{Related Work}

\textbf{Hate speech on social media.}
Detecting hate speech has been shown to be an important and  challenging task for society~\cite{ross2017measuring}.
Automatic hate speech detection methods are primarily text-based~\cite{lekea2018detecting, albadi2018they} or graph-based for hate speech diffusion~\cite{beatty2020graph}.
More recent methods use deep neural architectures like BERT to show superior performance than common models without overfitting \cite{nikolov2019nikolov}. 
Some studies are placed in the context of \covid{}~\cite{schild2020go, alshalan2020detection, vishwamitra2020analyzing}. However, none of the above-mentioned works detect and analyze the Anti-Asian hate and counter speech simultaneously over time during a pandemic, which is the gap we address in this work.

\textbf{Hate on social media during \covid{}.} 
Some studies to date have specifically addressed the spread of anti-Asian hate on social media~\cite{lu2020fear, vishwamitra2020analyzing}. \cite{schild2020go} proposed a multi-platform dataset of hate, but did not address counterspeech information, which is simultaneously prevalent on the platform. 
Recently, \cite{chen2020eyes} collected tweets containing controversial hashtags like \#ChineseVirus. However, as we have shown, the presence of hashtags is inaccurate to label a tweet as hate. We overcome this by developing a hand-labeled dataset and text-based classifier. 
Finally, contemporaneous work by \cite{vidgen2020detecting} released a large hand-labeled dataset of hate and counterspeech tweets. However, they do not conduct any analysis of the hate and counterspeech Twittersphere, which we present in this work, in addition to creating a complementary hand-labeled dataset.

\textbf{Counterspeech on social media.}
Counterspeech messaging has qualitatively been shown to be the most effective and the least intrusive solution to hate speech, though quantitative studies are limited \cite{gagliardone2015countering}. 
More recent work has focused on developing novel counterspeech data sets \cite{mathew2018analyzing}, detection classifiers \cite{mathew2019thou} and longitudinal analysis~\cite{garland2020countering}. 
But, none of these have studied Anti-Asian counterspeech in the context of \covid{}, which is the gap that we address in this work.

\section{Discussion and Conclusions}

Our findings in this work shed light on the important societal problem caused by the \covid{} pandemic. 
Notably, we observe that counterspeech might be effective in reducing the probability of neighbors becoming hateful. 
Our work paves the way towards the use of public counterspeech messaging campaigns as a potential solution against hate speech on social media.

Our work has a few limitations. 
First, our annotation scheme is coarse-grained (hate vs. counterspeech vs. neutral), which can be augmented with the recently released hand-labeled fine-grained dataset of \covid{}-related hateful and counterspeech tweets ~\cite{vidgen2020detecting} for deeper study, though the findings are not expected to change. We only study English language discussions. 
Our work also does not disentangle the influence of external events (e.g., Atlanta shooting). 
We will consider these in the  follow-up research.

\section*{Acknowledgements}
This research is supported in part by NSF (Expeditions CCF-1918770, NRT DGE-1545362, IIS-2027689), Adobe, Facebook, Microsoft, Georgia Institute of Technology, Russell Sage Foundation and the Institute for Data Engineering and Science (IDEAS) at Georgia Tech. We thank Manoj Niverthiand Haoran Zhang for help in annotation.

\bibliographystyle{IEEEtran}
\bibliography{main}

\begin{thebibliography}{10}
\providecommand{\url}[1]{#1}
\csname url@samestyle\endcsname
\providecommand{\newblock}{\relax}
\providecommand{\bibinfo}[2]{#2}
\providecommand{\BIBentrySTDinterwordspacing}{\spaceskip=0pt\relax}
\providecommand{\BIBentryALTinterwordstretchfactor}{4}
\providecommand{\BIBentryALTinterwordspacing}{\spaceskip=\fontdimen2\font plus
\BIBentryALTinterwordstretchfactor\fontdimen3\font minus
  \fontdimen4\font\relax}
\providecommand{\BIBforeignlanguage}[2]{{%
\expandafter\ifx\csname l@#1\endcsname\relax
\typeout{** WARNING: IEEEtran.bst: No hyphenation pattern has been}%
\typeout{** loaded for the language `#1'. Using the pattern for}%
\typeout{** the default language instead.}%
\else
\language=\csname l@#1\endcsname
\fi
#2}}
\providecommand{\BIBdecl}{\relax}
\BIBdecl

\bibitem{montemurro2020emotional}
N.~Montemurro, ``The emotional impact of covid-19: from medical staff to common
  people,'' \emph{Brain, behavior, and immunity}, 2020.

\bibitem{nbcnews}
\BIBentryALTinterwordspacing
{Kimmy Yam, NBC News}, ``Anti-asian hate incident reports nearly doubled in
  march, new data says,'' 2021, [Online; accessed 14-May-2021]. [Online].
  Available:
  \url{https://www.nbcnews.com/news/asian-america/anti-asian-hate-incident-reports}
\BIBentrySTDinterwordspacing

\bibitem{fbiracism}
\BIBentryALTinterwordspacing
J.~Margolin, \emph{FBI warns of potential surge in hate crimes against Asian
  Americans amid coronavirus (ABC News)}, 2020, last accessed: May 15, 2020.
  [Online]. Available:
  \url{https://abcnews.go.com/US/fbi-warns-potential-surge-hate-crimes-asian-americans/story?id=69831920}
\BIBentrySTDinterwordspacing

\bibitem{nyt}
\BIBentryALTinterwordspacing
{New York Times}, ``8 dead in atlanta spa shootings, with fears of anti-asian
  bias,'' 2021, [Online; accessed 14-May-2021]. [Online]. Available:
  \url{https://www.nytimes.com/live/2021/03/17/us/shooting-atlanta-acworth}
\BIBentrySTDinterwordspacing

\bibitem{saha2019prevalence}
K.~Saha, E.~Chandrasekharan, and M.~De~Choudhury, ``Prevalence and
  psychological effects of hateful speech in online college communities,'' in
  \emph{ACM WebSci}, 2019.

\bibitem{relia2019race}
K.~Relia, Z.~Li, S.~H. Cook, and R.~Chunara, ``Race, ethnicity and national
  origin-based discrimination in social media and hate crimes across 100 us
  cities,'' in \emph{ICWSM}.

\bibitem{vidgen2020detecting}
B.~Vidgen, A.~Botelho, D.~Broniatowski, E.~Guest, M.~Hall, H.~Margetts,
  R.~Tromble, Z.~Waseem, and S.~Hale, ``Detecting east asian prejudice on
  social media,'' \emph{arXiv preprint arXiv: 2005.03909}, 2020.

\bibitem{schild2020go}
F.~Tahmasbi, L.~Schild, C.~Ling, J.~Blackburn, G.~Stringhini, Y.~Zhang, and
  S.~Zannettou, ``" go eat a bat, chang!": On the emergence of sinophobic
  behavior on web communities in the face of covid-19,'' \emph{WWW}, 2021.

\bibitem{vishwamitra2020analyzing}
N.~Vishwamitra, R.~R. Hu, F.~Luo, L.~Cheng, M.~Costello, and Y.~Yang, ``On
  analyzing covid-19-related hate speech using bert attention,'' in
  \emph{ICMLA}.\hskip 1em plus 0.5em minus 0.4em\relax IEEE, 2020.

\bibitem{alshalan2020detection}
R.~Alshalan, H.~Al-Khalifa, D.~Alsaeed, H.~Al-Baity, and S.~Alshalan,
  ``Detection of hate speech in covid-19--related tweets in the arab region:
  Deep learning and topic modeling approach,'' \emph{Journal of Medical
  Internet Research}, vol.~22, no.~12, p. e22609, 2020.

\bibitem{al2021combating}
R.~Al-Jarf, ``Combating the covid-19 hate and racism speech on social media,''
  \emph{Technium Social Sciences Journal}, vol.~18, pp. 660--666, 2021.

\bibitem{lu2020fear}
R.~Lu and Y.~Sheng, ``From fear to hate: How the covid-19 pandemic sparks
  racial animus in the united states,'' \emph{Available at SSRN 3646880}, 2020.

\bibitem{chen2020covid}
E.~Chen, K.~Lerman, and E.~Ferrara, ``Covid-19: The first public coronavirus
  twitter dataset,'' \emph{arXiv preprint arXiv:2003.07372}, 2020.

\bibitem{dylan2020}
\BIBentryALTinterwordspacing
{Dylan Scott, VOX News}, ``Trump’s new fixation on using a racist name for
  the coronavirus is dangerous,'' 2020, [Online; accessed 14-May-2021].
  [Online]. Available:
  \url{https://www.vox.com/2020/3/18/21185478/coronavirus-usa-trump-chinese-virus}
\BIBentrySTDinterwordspacing

\bibitem{abc}
\BIBentryALTinterwordspacing
{Nydia Han, ABC News}, ``I don’t scare easily, but covid-19 virus of hate has
  me terrified: Reporter’s notebook,'' 2020, [Online; accessed 14-May-2021].
  [Online]. Available:
  \url{https://abcnews.go.com/US/asian-americans-covid-19-racism-virus-hate-reporters/story?id=70810109}
\BIBentrySTDinterwordspacing

\bibitem{businessinsider}
\BIBentryALTinterwordspacing
{Jeff Elder, Business Insider}, ``Activists fighting coronavirus-driven hate
  crimes are rallying on social media to turn masks into a symbol, rather than
  a target in racist attacks,'' 2020, [Online; accessed 14-May-2021]. [Online].
  Available:
  \url{https://www.businessinsider.com/mask-attack-covid19-racism-virus-hateisavirus-2020-4}
\BIBentrySTDinterwordspacing

\bibitem{prweek}
\BIBentryALTinterwordspacing
{Steve Barrett, PR Week}, ``Racism is a virus, not asians: Stopaapihate,''
  2020, [Online; accessed 14-May-2021]. [Online]. Available:
  \url{https://www.prweek.com/article/1711232/racism-virus-not-asians-stopaapihate}
\BIBentrySTDinterwordspacing

\bibitem{parekh2012there}
B.~Parekh \emph{et~al.}, ``Is there a case for banning hate speech?'' \emph{The
  content and context of hate speech: Rethinking regulation and responses}, pp.
  37--56, 2012.

\bibitem{fortuna2018survey}
P.~Fortuna and S.~Nunes, ``A survey on automatic detection of hate speech in
  text,'' \emph{ACM CSUR}, vol.~51, no.~4, pp. 1--30, 2018.

\bibitem{roberto2020stigmatization}
K.~J. Roberto, A.~F. Johnson, and B.~M. Rauhaus, ``Stigmatization and prejudice
  during the covid-19 pandemic,'' \emph{Administrative Theory \& Praxis},
  vol.~42, no.~3, pp. 364--378, 2020.

\bibitem{reny2020xenophobia}
T.~T. Reny and M.~A. Barreto, ``Xenophobia in the time of pandemic: othering,
  anti-asian attitudes, and covid-19,'' \emph{Politics, Groups, and
  Identities}, pp. 1--24, 2020.

\bibitem{gover2020anti}
A.~R. Gover, S.~B. Harper, and L.~Langton, ``Anti-asian hate crime during the
  covid-19 pandemic: Exploring the reproduction of inequality,'' \emph{American
  journal of criminal justice}, vol.~45, no.~4, pp. 647--667, 2020.

\bibitem{kam2019infectious}
C.~D. Kam, ``Infectious disease, disgust, and imagining the other,'' \emph{The
  Journal of Politics}, vol.~81, no.~4, pp. 1371--1387, 2019.

\bibitem{ross2017measuring}
B.~Ross, M.~Rist, G.~Carbonell, B.~Cabrera, N.~Kurowsky, and M.~Wojatzki,
  ``Measuring the reliability of hate speech annotations: The case of the
  european refugee crisis,'' \emph{arXiv preprint arXiv:1701.08118}, 2017.

\bibitem{devlin2018bert}
J.~Devlin, M.-W. Chang, K.~Lee, and K.~Toutanova, ``Bert: Pre-training of deep
  bidirectional transformers for language understanding,'' \emph{arXiv preprint
  arXiv:1810.04805}, 2018.

\bibitem{leskovec2010signed}
J.~Leskovec, D.~Huttenlocher, and J.~Kleinberg, ``Signed networks in social
  media,'' in \emph{Proceedings of the SIGCHI conference on human factors in
  computing systems}, 2010, pp. 1361--1370.

\bibitem{del2016echo}
M.~Del~Vicario, G.~Vivaldo, A.~Bessi, F.~Zollo, A.~Scala, G.~Caldarelli, and
  W.~Quattrociocchi, ``Echo chambers: Emotional contagion and group
  polarization on facebook,'' \emph{Scientific reports}, vol.~6, p. 37825,
  2016.

\bibitem{mathew2019spread}
B.~Mathew, R.~Dutt, P.~Goyal, and A.~Mukherjee, ``Spread of hate speech in
  online social media,'' in \emph{ACM WebSci}, 2019, pp. 173--182.

\bibitem{anagnostopoulos2008influence}
A.~Anagnostopoulos, R.~Kumar, and M.~Mahdian, ``Influence and correlation in
  social networks,'' in \emph{ACM SIGKDD}, 2008.

\bibitem{lekea2018detecting}
I.~K. Lekea and P.~Karampelas, ``Detecting hate speech within the terrorist
  argument: a greek case,'' in \emph{ASONAM}.\hskip 1em plus 0.5em minus
  0.4em\relax IEEE, 2018, pp. 1084--1091.

\bibitem{albadi2018they}
N.~Albadi, M.~Kurdi, and S.~Mishra, ``Are they our brothers? analysis and
  detection of religious hate speech in the arabic twittersphere,'' in
  \emph{ASONAM}.\hskip 1em plus 0.5em minus 0.4em\relax IEEE, 2018, pp. 69--76.

\bibitem{beatty2020graph}
M.~Beatty, ``Graph-based methods to detect hate speech diffusion on twitter,''
  in \emph{ASONAM}.\hskip 1em plus 0.5em minus 0.4em\relax IEEE, 2020, pp.
  502--506.

\bibitem{nikolov2019nikolov}
A.~Nikolov and V.~Radivchev, ``Nikolov-radivchev at semeval-2019 task 6:
  Offensive tweet classification with bert and ensembles,'' in \emph{IWSE},
  2019.

\bibitem{chen2020eyes}
L.~Chen, H.~Lyu, T.~Yang, Y.~Wang, and J.~Luo, ``In the eyes of the beholder:
  Sentiment and topic analyses on social media use of neutral and controversial
  terms for covid-19,'' \emph{arXiv preprint arXiv:2004.10225}, 2020.

\bibitem{gagliardone2015countering}
I.~Gagliardone, D.~Gal, T.~Alves, and G.~Martinez, \emph{Countering online hate
  speech}, 2015.

\bibitem{mathew2018analyzing}
B.~Mathew, N.~Kumar, P.~Goyal, A.~Mukherjee \emph{et~al.}, ``Analyzing the hate
  and counter speech accounts on twitter,'' \emph{arXiv preprint
  arXiv:1812.02712}, 2018.

\bibitem{mathew2019thou}
B.~Mathew, P.~Saha, H.~Tharad, S.~Rajgaria, P.~Singhania, S.~K. Maity,
  P.~Goyal, and A.~Mukherjee, ``Thou shalt not hate: Countering online hate
  speech,'' in \emph{ICWSM}, vol.~13, no.~01, 2019, pp. 369--380.

\bibitem{garland2020countering}
J.~Garland, K.~Ghazi-Zahedi, J.-G. Young, L.~H{\'e}bert-Dufresne, and
  M.~Galesic, ``Countering hate on social media: Large scale classification of
  hate and counter speech,'' \emph{arXiv preprint arXiv:2006.01974}, 2020.

\end{thebibliography}

\end{document}